\documentclass{article}
\pdfoutput=1
\usepackage{times}
\usepackage{pstricks}
\usepackage{graphicx,color}
\DeclareGraphicsExtensions{.jpg,.jpeg,.pdf,.mps,.png,.pdftex}
\usepackage{graphics,latexsym,amsmath,amssymb,xspace,enumerate,paralist}
\usepackage{wrapfig,subfigure}
\usepackage[noend]{algorithmic}
\usepackage{algorithm}
\usepackage{amsthm}
\usepackage{setspace}
\usepackage[left=2.5cm,top=2cm,right=2cm,bottom=2.5cm]{geometry}

\newcounter{pseudocode}

\newtheorem{theorem}{Theorem}[section]
\newtheorem{observation}{Observation}[section]
\newtheorem{definition}[theorem]{Definition}

\long\def\mc{MapConcatenate\xspace}

\newcommand {\comment}[1]{}

\newlength\colwidth

\newcount\hours
\newcount\minutes
\def\SetTime{\hours=\time
        \global\divide\hours by 60
        \minutes=\hours
        \multiply\minutes by 60
        \advance\minutes by-\time
        \global\multiply\minutes by-1 
        \ifnum\hours<12 \def\ampm{am} 
        \ifnum\hours<1 \advance\hours by+12 \fi
        \else
        \def\ampm{pm} \advance\hours by-12 \fi }
\SetTime

\def\now{\number\hours:\ifnum\minutes<10 0\fi\number\minutes\ampm}

\makeatletter
\def\cramped                                    
  {\parskip\@outerparskip\@topsep\parskip       
  \@topsepadd2pt\itemsep0pt                     %
}
\makeatother

\begin{document}
\title{Towards Chip-on-Chip Neuroscience: \\Fast Mining of Frequent 
Episodes Using Graphics Processors}

\author{Yong Cao, Debprakash Patnaik, Sean Ponce, Jeremy Archuleta,\\ [0.2ex]
Patrick Butler, Wu-chun Feng, and Naren Ramakrishnan\\ [0.2ex]
Department of Computer Science\\ [0.2ex]
Virginia Tech, VA 24061, USA}

\singlespace
\date{} \maketitle

\doublespacing

\begin{abstract}
\noindent

Computational neuroscience is being revolutionized with the advent of multi-electrode arrays that provide real-time, dynamic, perspectives into brain function. Mining event streams from these chips is critical to understanding the firing patterns of neurons and to gaining insight into the underlying cellular activity. We present a GPGPU solution to mining spike trains. We focus on mining frequent episodes which captures coordinated events across time even in the presence of intervening background/``junk'' events. Our algorithmic contributions are two-fold: MapConcatenate, a new computation-to-core mapping scheme, and a two-pass elimination approach to quickly find supported episodes from a large number of candidates. Together, they help realize a real-time ``chip-on-chip'' solution to neuroscience data mining, where one chip (the multi-electrode array) supplies the spike train data and another (the GPGPU) mines it at a scale unachievable previously. Evaluation on both synthetic and real datasets demonstrate the potential of our approach.

\end{abstract}


\section{Introduction}
\label{intro}
Temporal (symbolic) event streams are popular in scientific domains, such as 
physical plants, medical diagnostics, and neuroscience. 
In all these domains, we are given occurrences of events of interest over a time course
and the goal is to identify trends and behaviors that serve discriminatory or
descriptive purposes.
In this paper, we focus exclusively on event streams gathered through
multi-electrode array (MEA) chips for studying neuronal function although our algorithms
and implementations are applicable to a wider variety of domains.

\begin{figure}
\centering
\begin{tabular}{cc}
\includegraphics[width=3in]{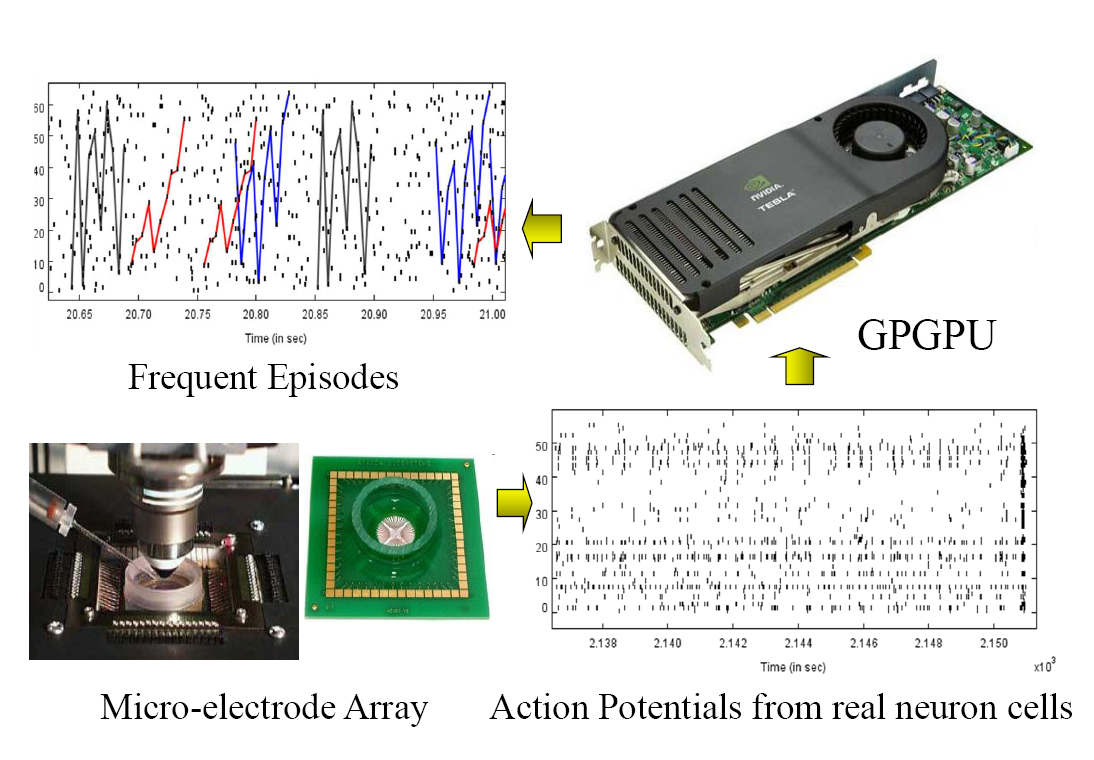} &
\includegraphics[width=3in]{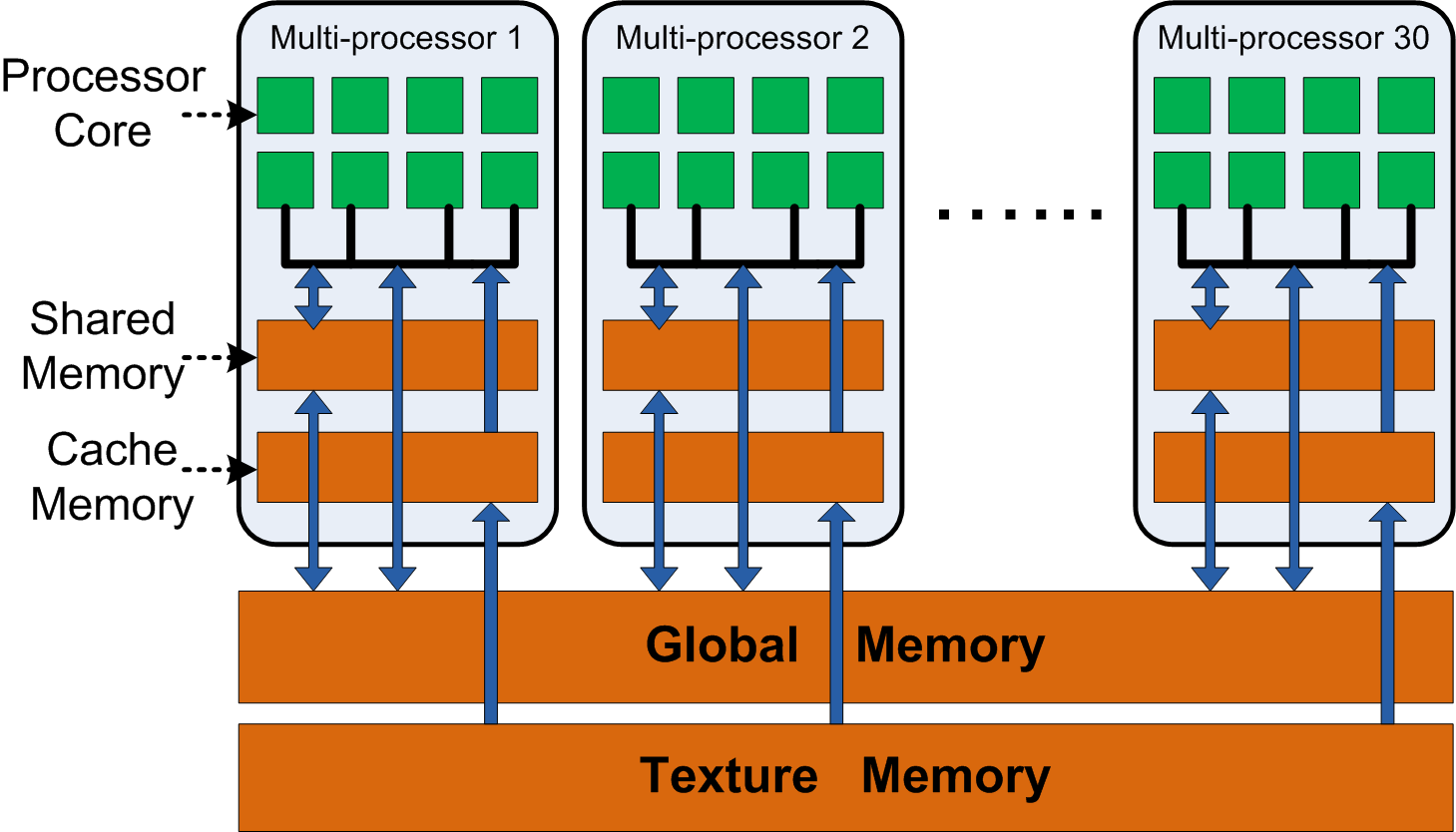} \\
\end{tabular}
\caption{(left) Chip-on-Chip neuroscience: spike trains recorded from a multi-electrode array (MEA) are mined by 
a GPGPU to yield frequent episodes which can be summarized to reconstruct the underlying neuronal circuitry. (right) Architecture of the NVIDIA GTX280 GPGPU in
detail.}
\label{fig:chip}
\end{figure}

An MEA records spiking action potentials from an ensemble of neurons which after
various pre-processing steps, yields a 
spike train dataset providing real-time, dynamic, perspectives into
brain function (see Figure~\ref{fig:chip}). 
Identifying sequences (e.g., cascades) 
of firing neurons, determining their characteristic delays, and reconstructing the functional
connectivity of neuronal circuits are key problems of interest. This provides
critical insights into the cellular activity recorded in the neuronal tissue.

With rapid developments in instrumentation and data acquisition
technology, the size of event streams recorded by MEAs has
concomitantly grown, leading us to exhaust the abilities of sequential computation.
For instance, just a few minutes of cortical
recording from a 64-channel MEA can easily generate millions of
spikes!  It has thus become imperative to enable fast, near real-time
computation, data mining, and visualization of mined patterns.  
This mapping is critical as it can result in orders-of-magnitude difference in performance.

%
%
%
We adopt GPGPUs as the platform of choice for neuroscience
data mining for several reasons. First, they are particularly
economical in morphing a desktop into a rather powerful parallel
computing machine (price-to-performance ratio). This is especially
attractive to neuroscientists who might not have access to clusters of
workstations.
%
%
At the same time, mapping our application to a GPGPU architecture is
non-trivial. GPGPUs were originally designed for data-parallel
applications (e.g., rendering), not quite the class of problems that
temporal data mining falls under.
Our main contributions are as follows:

\begin{enumerate}
\item {\it {\mc}}, a computation-to-core mapping scheme suitable for
frequent episode mining that takes advantage of the computational power of hundreds of
processing cores on a GPU. {\it {\mc}} is a two-stage scheme like MapReduce~\cite{Dean:2004}
but quite different in the intent and applicability of both the stages.
\item A two-pass elimination approach to find supported temporal episodes from a large 
number of candidates. The first pass, which is conducted with relaxed timing constraints, can eliminate
most of the non-supported episode candidates with the accurate count computed
using the second pass. Design of this elimination approach and proving
its correctness is non-trivial since the set of episodes returned by the relaxed set 
of constraints is {\bf not} a superset of the original set of constraints but
the counts are indeed an upper bound. 
\item A `chip-on-chip' real-time solution to neuroscience data mining where
one chip (the MEA)
provides
the data and another chip (the graphics processor) mines it. Our solution is
not a complete data streaming solution; nevertheless,
we achieve real-time responsiveness by processing partitions of the data stream in turn.
\end{enumerate}
%

\section{Problem Statement}
A spike train dataset can be modeled as an event stream, where
each symbol/event type corresponds to a specific neuron (or clump of neurons) and the dataset encodes
occurrence times of these events over the time course.

\begin{definition}
An event stream is denoted by a sequence of events $\langle(E_1, t_1),(E_2, t_2),\ldots(E_n, t_n)\rangle$, where $n$ is the total number of events. Each event $(E_i, t_i)$ is characterized by an event type $E_i$ and a time of occurrence $t_i$. The sequence is ordered by time i.e. $t_i\leq t_{i+1} \forall i = 1,\ldots,n-1$ and $E_i$'s are drawn from a finite set $\xi$.
\end{definition}

One class of interesting patterns that we wish to
 discover are frequently occurring groups of events (i.e., firing cascades of
neurons) with some constraints on 
 ordering and timing of these events. This is captured by the notion of episodes,
the original framework for which was proposed by Mannila \textit{et al}~\cite{window}.

\begin{definition}
An (serial) episode $\alpha$ is an ordered
tuple of event types $V_{\alpha} \subseteq \xi$.
\end{definition}

For example $(A \rightarrow B \rightarrow C \rightarrow D)$ is a 4-node serial episode, and it
captures the pattern that
neuron A fires, followed by neurons B, C and D in that order, but not necessarily without intervening `junk' firings of
neurons (even possibly neurons A, B, C, or D).
This ability to intersperse noise or don't care states, of arbitrary length,
between the event symbols in the definition of an episode is what makes these patterns
non-trivial, useful, and 
comprehensible.

\textit{Frequency of episodes}: 
The notion of frequency of an episode
can 
be defined in several ways. 
In \cite{window}, it is defined as the fraction of windows in which the episode occurs. Another measure of frequency is the non-overlapped count which is the size of the largest set of non-overlapped occurrences of an episode. Two occurrences are non-overlapped if no event of one occurrence appears in between the events of the other. 
In the event stream of Fig.~\ref{fig:counter-eg}, there are at most two non-overlapped
occurrences of the episode $A \rightarrow B$, although there are 8 occurrences in total.

We use the non-overlapped occurrence count as the frequency measure of choice
due to its strong theoretical properties under
a generative model of events~\cite{vatsan2}.
It has also been
argued in \cite{Patnaik2008} that, for the neuroscience application, counting non-overlapped occurrences is 
natural because episodes then correspond to
causative, ``syn-fire'', chains that happen at different times again and again.



\textit{Temporal constraints}: Besides the frequency threshold, a further level of 
constraint can be imposed on the definition of episodes. In multi-neuronal datasets,
if one would like to infer that neuron $A$'s firings cause a neuron $B$ to fire, then spikes from neuron $B$ 
cannot occur immediately or spontaneously after $A$'s spikes due to axonal
conduction delays.
These spikes cannot also occur too much later than $A$ for the same reason.
Such minimum and maximum inter-event delays are common in
other application domains as well.
We hence place inter-event time constraints giving rise to
episodes such as:
\begin{equation*}
(A^{\underrightarrow{(t_{low}^{1},t_{high}^{1}]}}B^{\underrightarrow{(t_{low}^{2},t_{high}^{2}]}}C)
\label{eq:episode-with-interval}
\end{equation*}
In a given occurrence of episode $A\rightarrow$ $B\rightarrow$ $C$ 
let $t_{A}$, $t_{B}$, and $t_{C}$ denote the time of occurrence of corresponding event types. A valid occurrence of the serial episode satisfies $$t_{low}^{1} < (t_{B}-t_{A}) \le t_{high}^{1},
t_{low}^{2} < (t_{C}-t_{B}) \le t_{high}^{2}$$
(In general, an $N$-node serial episode is associated with $N-1$ inter-event constraints.)
In Fig.~\ref{fig:counter-eg}, there is exactly one occurrence of the episode 
$A\stackrel{(5,10]}{\rightarrow}B\stackrel{(10,15]}{\rightarrow}C$
satisfying the desired inter-event constraints (shown in dotted boxes).

\begin{quote}
{\it Problem 1:} Given an event stream $\{(E_i, t_i)\}$, $i\in\{1\ldots{n}\}$, a set of inter-event constraints $I=\{(t_{low}^k,t_{high}^k]\}$,$k\in\{1\ldots{l}\}$,\, find all serial episodes $\alpha$ of the form:
\begin{equation}
\alpha = \langle E_{(1)}^{\underrightarrow{(t_{low}^{(1)},t_{high}^{(1)}]}} E_{(2)} \ldots^{\underrightarrow{(t_{low}^{(N-1)},t_{high}^{(N-1)}]}} E_{(N)} \rangle \nonumber
\end{equation}
such that the non-overlapped occurrence counts of each episode $\alpha$ is $\geq \theta$, a user-specified
threshold.
Here $E_{(.)}$'s are the event types in the episode $\alpha$ and $(t_{low}^{(.)},t_{high}^{(.)}]$'s $\in I$ are the corresponding inter-event constraints. 
\label{def:problem}
\end{quote}

\section{Prior Work}
We review prior work in three categories: mining frequent episodes,
data mining using GPGPUs, and the map-reduce framework for large scale
computations.

\noindent
{\bf Mining Frequent Episodes:} The overall mining procedure for frequent episodes is based on 
level-wise mining. Within this framework there are two
broad classes of algorithms: window-based~\cite{window} and state machine
based~\cite{fsa-mining,vatsan2}, and they primarily differ in how
they define the frequency of an episode. 
The window based algorithms define frequency of an episode as the
fraction of windows on the event sequence in which the episode occurs.
The state machine based algorithms are more efficient and define frequency as the size of largest set of non-overlapped occurrences of an episode.
Within the class of state machine algorithms,
serial episode discovery using non-overlapped counts was described
in~\cite{vatsan2}, and their extension to temporal constraints is
given in~\cite{Patnaik2008}. With the introduction of temporal constraints 
the state machine based algorithms become more complicated.  
They must keep track of what part of an episode has been seen, 
which event is expected next and, when episodes inter-leave, they 
must make a decision of which events to be used in the formation 
of an episode.

%
%

\noindent
{\bf Data Mining Using GPGPUs:} Many researchers have harnessed the capabilities of GPGPUs for data mining.
The key to porting a data mining algorithm onto a GPGPU is to, in one sense, ``dumb it down'';
i.e., conditionals, program branches, and complex decision constraints are not
easily parallelizable on a GPGPU and algorithms using these constraints
will require significant reworking to fit this architecture (temporal episode
mining unfortunately falls in this category). There are many emerging publications
in this area but due to space restrictions, we survey only a few here.
The SIGKDD tutorial by Guha et al.~\cite{tutorial} provides a gentle introduction to the aspects of data mining on GPGPUs through problems like
k-means clustering, reverse nearest neighbor(RNN), discrete wavelet transform(DWT), sorting networks, etc.
In~\cite{dmg2}, a bitmap technique is proposed to support counting and is used
to design GPGPU variants of {\it Apriori} and k-means clustering. This work
also proposes co-processing for itemset mining where the complicated tie data
structure is kept and updated in the main memory of CPU and only the
itemset counting is executed in parallel on GPU.
A sorting algorithm on GPGPUs with applications to frequency counting and histogram construction
is discussed in~\cite{dmg1} which essentially recreates sorting networks on the GPU.
Li et al.~\cite{li-cut} present a `cut-and-stitch' algorithm for approximate learning
of Kalman filters. Although this is not a GPGPU solution {\it per se}, we point out
that our proposed approach shares with this work the difficulties of mining temporal behavior 
in a parallel context. 

\noindent
{\bf MapReduce:} Modeled after LISP primitives,
MapReduce~\cite{Dean:2004} provides a distribution framework for large scale computations using two functions:
{\it map}, and {\it reduce}. 
It has received a fair share of attention (and, sometimes, criticism) from the
cloud computing and data-intensive computing communities. 
The framework has been ported to many platforms, including
GPGPUs (e.g., see~\cite{mars}). See Section~\ref{sec:MapConcat} 
for comparisons between our proposed framework and MapReduce.

\section{GPGPU Architecture}
To understand the algorithmic details behind our approach, we 
provide a brief overview of GPGPU and its architecture.

The initial purpose of specialized GPUs was to accelerate the display
of 2D and 3D graphics, much in the same way that the FPU focused on
accelerating floating-point instructions.  However, the rapid
technological advances of the GPU, coupled with extraordinary
speed-ups of application ``toy'' benchmarks on the specialized GPUs,
led GPU vendors to transform the GPU from a specialized processor to a
general-purpose graphics processing unit (GPGPU), such as the NVIDIA GTX
280, as shown in Figure~\ref{fig:chip} (right).  To lessen the steep learning
curve, GPU vendors also introduced programming environments, such as
the Compute Unified Device Architecture (CUDA).

%

\noindent
{\bf Processing Elements:}
The basic execution unit on the GTX 280 is a scalar processing
\textbf{core}, of which 8 together form a
\textbf{multiprocessor}. While the number of multiprocessors and
processor clock frequency depends on the make and model of the GPU,
every multiprocessor in CUDA executes in SIMT (Single Instruction,
Multiple Thread) fashion, which is similar in nature to SIMD (Single
Instruction, Multiple Data) execution. Specifically, groups of 32
threads form a \textbf{warp} and execute the exact same codepath until
every thread terminates. However, when codepaths diverge, each thread
must now execute every instruction on \emph{every} thread path,
therefore, implying that optimal performance is attained when all 32
threads do \emph{not} branch down different codepaths. The execution
of a single instruction with 8 cores and a warp size of 32 completes
in 4 cycles.

\noindent
{\bf Memory Hierarchy:}
The GTX 280 contains multiple forms of memory. Beginning with the
furthest from the GPU processing elements, the \textbf{device memory}
is located off-chip on the graphics card and provides the main source
of storage for the GPU while simultaneously being accessible from the
CPU and GPU. Each multiprocessor on the GPU contains three caches ---
a \textbf{texture cache}, \textbf{constant cache}, and \textbf{shared
  memory}. The texture cache and constant cache are both
\emph{read-only} memory providing fast access to immutable
data. Shared memory, on the other hand, is \emph{read-write} to
provide each core with fast access to the shared address space of a
thread block within a multiprocessor. Finally, on each core resides a plethora of registers
such that there exists minimal reliance on local memory resident
off-chip on the device memory.

\noindent
{\bf Parallelism Abstractions:}
At the highest level, the CUDA programming model requires the
programmer to offload functionality to the GPGPU as a compute
\textbf{kernel}. This kernel is evaluated as a set of \textbf{thread
  blocks} logically arranged in a \textbf{grid} to facilitate
organization. In turn, each block contains a set of \textbf{threads},
which will be executed on the same multiprocessor, with the threads
scheduled in warps, as mentioned above.

\section{Algorithms}
\label{sec:algorithms}
Our overall approach for solving Problem 1 is based on a state machine algorithm with 
inter-event constraints~\cite{Patnaik2008}. 
There are two major phases of this algorithm: generating episode candidates and counting these episodes, 
and we focus on the latter since it is 
the key performance bottleneck, typically by a few orders of magnitude.
Therefore, we present a GPU-accelerated algorithm for counting episodes 
of various size, while candidate generation is executed sequentially on
a CPU.

In this section, we first present the standard sequential algorithm for mining frequent episodes
with inter-event constraints.
Next, we introduce our GPU based algorithm (called $A1$). We then propose a two-pass counting approach, where the first counting pass eliminates most of unsupported episodes and the second counting pass completes the 
counting tasks for the remaining episodes. Since the first counting pass uses a less complex algorithm (called
$A2$), the execution time saved at this step contributes to an overall performance gain even we 
go through two-pass counting ($A2 + A1$).

\subsection{Episode Mining with Inter-Event Constraints}
Algorithm \ref{alg:A1} outlines the serial counting procedure for a single 
episode $\alpha$. 
%
Briefly, it maintains a data structure $s$ which is a list of lists. Each list $s[k]$ in $s$ corresponds to an event type $E_{(k)}\in \alpha$ and stores the times of occurrences of those events with event-type $E_{(k)}$ which satisfy the inter-event constraint $(t_{low}^{(k-1)},t_{high}^{(k-1)}]$ with at least one entry $t_{j}\in s[k-1]$. This requirement is relaxed for $s[0]$, thus every time an event $E_{(0)}$ is seen in the data its occurrence time is pushed into $s[0]$.

\begin{figure}[htbp]
	\centering
		\includegraphics[width=2.0in]{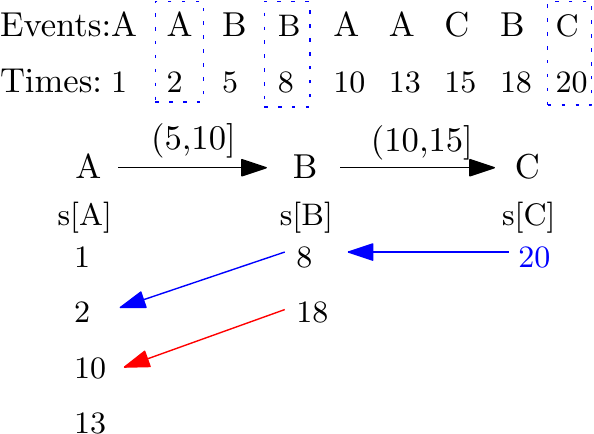}
	\caption{Illustration of the data structure $s$ for counting $A\stackrel{(5,10]}{\rightarrow}B\stackrel{(10,15]}{\rightarrow}C$}
	\label{fig:counter-eg}
\end{figure}

When an event of type $E_{(k)}, 2\leq{k}\leq{N}$ at time $t$ is seen, we look for an entry $t_j \in s[k-1]$ such that $t - t_j \in (t_{low}^{(k-1)},t_{high}^{(k-1)}]$. Therefore, if we are able to add the event to the list $s[k]$, it implies that there exists at least one previous event with event-type $E_{(k-1)}$  in the data stream for the current event which satisfies the inter-event constraint between $E_{(k-1)}$ and $E_{(k)}$. Applying this argument recursively, if we are able to add an event with event-type $E_{(|\alpha|)}$ to its corresponding list in $s$, then there exists a sequence of events corresponding to each event type in $\alpha$ satisfying the respective inter-event constraints. Such an event marks the end of an occurrence after which the $count$ for $\alpha$ is incremented and the data structure $s$ is reinitialized. Figure \ref{fig:counter-eg} illustrates the data structure $s$ for counting $A\stackrel{(5,10]}{\rightarrow}B\stackrel{(10,15]}{\rightarrow}C$. 

The maximality of the left-most and inner-most occurrence counts for a general serial episode has been shown in \cite{vatsan2}. Similar arguments hold for the case of serial episodes with inter-event constraints and not shown here for lack of space.

\singlespace

\begin{algorithm}[ht]
\begin{algorithmic}[1]
\REQUIRE Candidate $N$-node episode $\alpha=\langle E_{(1)}\stackrel{(t_{low}^{(1)},t_{high}^{(1)}]}{\longrightarrow}\ldots E_{(N)}\rangle$ and event sequence $\textsl{S}=\{(E_i, t_i)\}, i \in \{1\ldots{n}\}$.
\ENSURE Count of non-overlapped occurrences of $\alpha$ satisfying inter-event constraints
\STATE $count = 0$; $s=[[],\ldots,[]]$ //List of $|\alpha|$ lists
\FORALL{$(E,t) \in \textsl{S}$}
	\FOR{$i = |\alpha| \mbox{ to } 1$} \label{line:outer}
		\STATE $E_{(i)} = i^{th}$ event type $\in \alpha$
		\IF{$E = E_{(i)}$}
			\STATE $i_{prev} = i - 1$
			\IF{$i > 1$}
				\STATE $k = |s[i_{prev}]|$
				\WHILE{$k > 0$}\label{line:inner}
					\STATE $t_{prev} = s[i_{prev},k]$
					\IF{$t_{low}^{(i_{prev})} < t - t_{prev} \leq t_{high}^{(i_{prev})}$}
						\IF{$i = |\alpha|-1$}
							\STATE $count ++$; $s=[[],\ldots,[]]$; \textbf{break} Line: \ref{line:outer}
						\ELSE
							\STATE $s[i]=s[i]\cup{t}$
						\ENDIF
						\STATE \textbf{break} Line: \ref{line:inner}
					\ENDIF
					\STATE $k = k - 1$
				\ENDWHILE
			\ELSE 
				\STATE $s[i]=s[i]\cup{t}$
			\ENDIF
		\ENDIF
	\ENDFOR
\ENDFOR
\STATE RETURN count
\end{algorithmic}
\caption{Serial Episode Mining}
\label{alg:A1}
\end{algorithm}

\doublespace





\subsection{GPU Mining Algorithm: A1}  

\subsubsection{PTPE: per-thread per-episode}

In order to design a parallely-executed algorithm for counting supports for
$M$ episodes, we need to identify a set of independent computational tasks/units that can be mapped onto different computational cores of GPU. A standard approach is to
use $M$
computational units, each of which is counting for one episode. We then map each computational unit to a logic GPU thread, which can execute in parallel with other threads. Because the counting of one episode in each computational unit is executed sequentially, like CPU, we simply implement Algorithm~\ref{alg:A1} as the CUDA kernel for each counting thread.



\subsubsection{\mc: multiple threads per episode}
\label{sec:MapConcat}
When the number of episodes is smaller than the number of cores, a per-thread per-episode model is prone to inefficiencies in core usage , leading to higher execution times.
Hence we seek to achieve a higher level of parallelism within the counting of a single episode.
We segment the input event stream into segments of sub-streams and use each thread block to count one segment. 
The subcounts of all segements
need to be concatenated together for the final count, and hence
we refer to this two-step approach as ``{\it \mc}'', which include a {\em Map} step, where each event segment is mapped to one thread block for local counting, and a {\em Concatenate} step, where the counting results from 
neighboring segments coresponding to episodes that straddle them are concatenated/merged together.

\textbf{{\em MapReduce vs \mc}} 

This new computation-to-core mapping scheme, {\it \mc}, is quite different from 
MapReduce~\cite{Dean:2004}, in both stages.
Whereas the {\it map} functions in MapReduce operate independently across segments,
the {\it map} functions in {\mc} require data access from adjacent segments. 
Similarly, whereas the results from any two segments can be ``reduced'' in MapReduce,
only the results from two adjacent segments can be concatenated in \mc.

%


\textbf{{\em Design of  \mc}} 

When we divide the input stream into segments, there are chances that some occurrences of an episode 
span across the boundaries of consecutive segments. 
\begin{figure}
	\centering
		\includegraphics[width=3.8in]{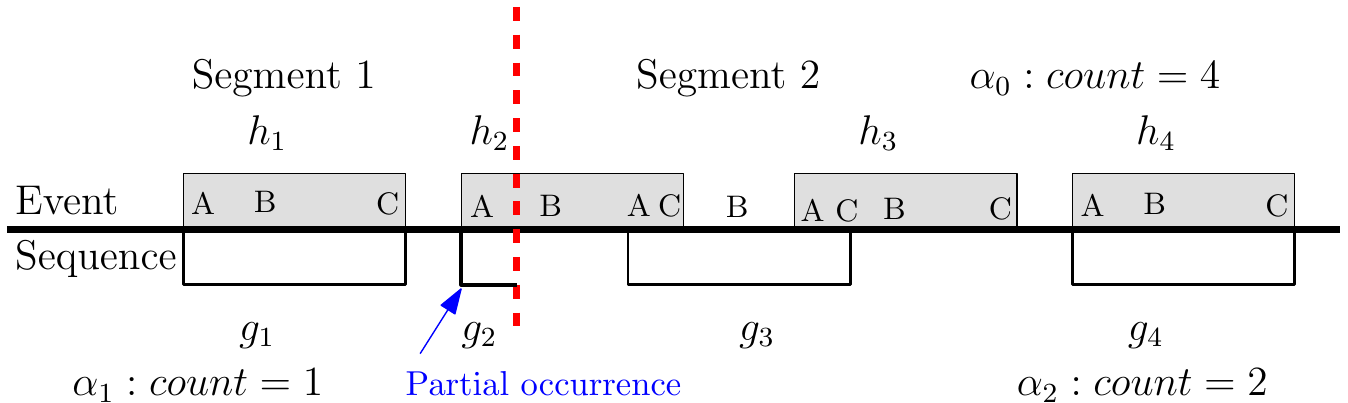}
	\caption{Illustration of splitting a data sequence into segments and counting within each segment.}
	\label{fig:map-eg}
\end{figure}
As an example, see Fig.~\ref{fig:map-eg} which depicts
a data sequence divided into two segments. The shaded rectangles on the top mark the non-overlapped occurrences $h_1\ldots h_4$ of an episode 
$(A \rightarrow B \rightarrow C)$ (assume for now
that inter-event constraints are always satisfied), as seen by a state machine $\alpha_0$ on the unified event sequence. $\alpha_0$ is thus the reference (serial) state machine.
Let $\alpha_1$ and $\alpha_2$ be the state machines counting occurrences in segment 1 and segment 2 respectively. During the {\em Map} step, $\alpha_1$ and $\alpha_2$ are executed in parallel, and each state machine can see a local view of the episode occurrences, which are shown by empty rectangles below the event sequence.  
$\alpha_1$ sees the occurrence $g_1$ and a partial occurrence $g_2$. For $\alpha_2$, it will first see the occurrence $g_3$ and therefore miss $h_3$ before moving onto $g_4$. 


We propose 
{\em Map} and {\em Concatenate} steps that use {\it multiple state machines} in each segment, so 
that the counting of a segment is able to anticipate partial occurrences near boundaries.
Let us explain in detail why multiple state machines are necessary, and how we design the {\em Map} step and {\em Concatenate} step to maintain the correctness of counting.

\begin{figure}[ht]
	\centering
		\includegraphics[width=2.4in]{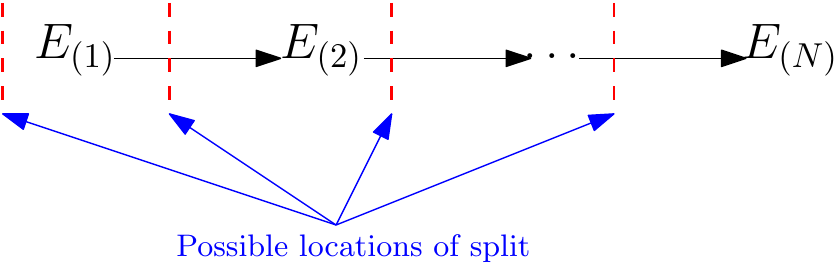}
	\caption{Illustration of different possibilities of an occurrence splitting across two adjacent data segments.}
	\label{fig:eps-split}
\end{figure}

Assume that we are counting an episode $\alpha=\langle E_{(1)}\stackrel{(t_{low}^{(1)},t_{high}^{(1)}]}{\longrightarrow}\ldots E_{(N)} \rangle$, the data sequence is divided into $P$ segments,
and events in the $p^{th}$ data segment are in the range $(\tau_p, \tau_{p+1}]$. An occurrence of episode $\alpha$ can be split across two adjacent segments in at least $N$ ways as shown in Figure~\ref{fig:eps-split}.
For each possible split, we need one state machine,  $\alpha_p^k, 0 \leq k \leq N-1$, to count the second segment, 
starting at $t = \tau_p - \sum_{i=1}^{k}{t_{high}^{(i)}}$. So we have $N$ different state machines all counting occurrences of episode $alpha$ using Algorithm~\ref{alg:A1}, handling all possible cases of split between current segment and previous segment.

\begin{figure}[ht]
	\centering
		\includegraphics[width=3.4in]{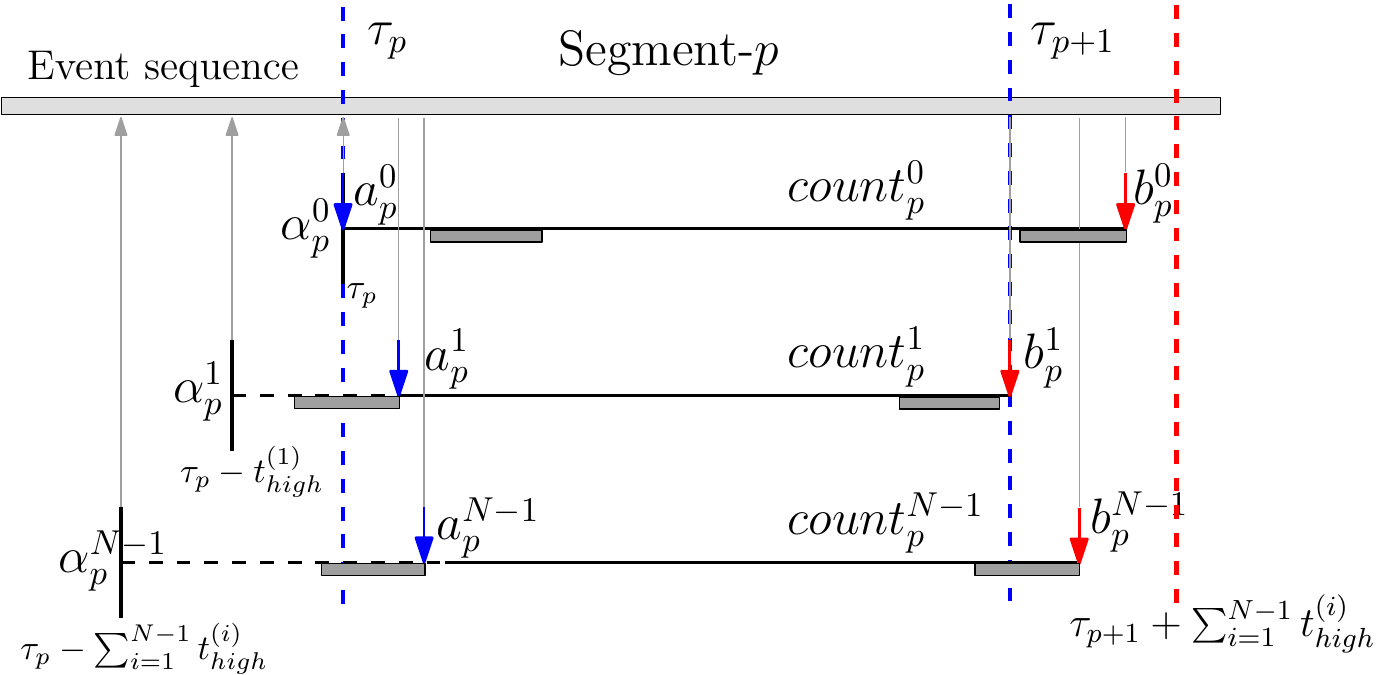}
	\caption{Illustration of a {\em Map} step.}
	\label{fig:map-step}
\end{figure}

For each segment $p$, the {\em Map} step is designed as follows, and illustrated in Figure~\ref{fig:map-step}.
\begin{enumerate}
\setlength{\itemsep}{0pt}
	\item Each state machine $\alpha_p^k$ maintains its own count $= count_p^k$.
	\item $\alpha_p^k$ does not increment count for occurrences ending at time $t \leq \tau_p$.
	\item $\alpha_p^k$ stores the end time of the first occurrence that completes at time $t$, $\tau_p < t < \tau_p + \sum_{i=1}^{N-1}{t_{high}^{(i)}}$. Let this be $a^k_p$. If there is no such occurrence $a^k_p$ is set to $\tau_p$.
	\item $\alpha_p^k$ on reaching end of the segment, crosses over into the next segment to completes the current partial occurrence and continues until $t < \tau_{p+1}+\sum_{i=1}^{N-1}{t_{high}^{(i)}}$. Let the end time of this occurrence be $b^k_p$. Note that count is not incremented for this occurrence.
	In case the occurrence cannot be completed $b^k_p$ is set to $\tau_{p+1}$.
\end{enumerate}

\begin{figure}[ht]
	\centering
		\includegraphics[width=4.6in]{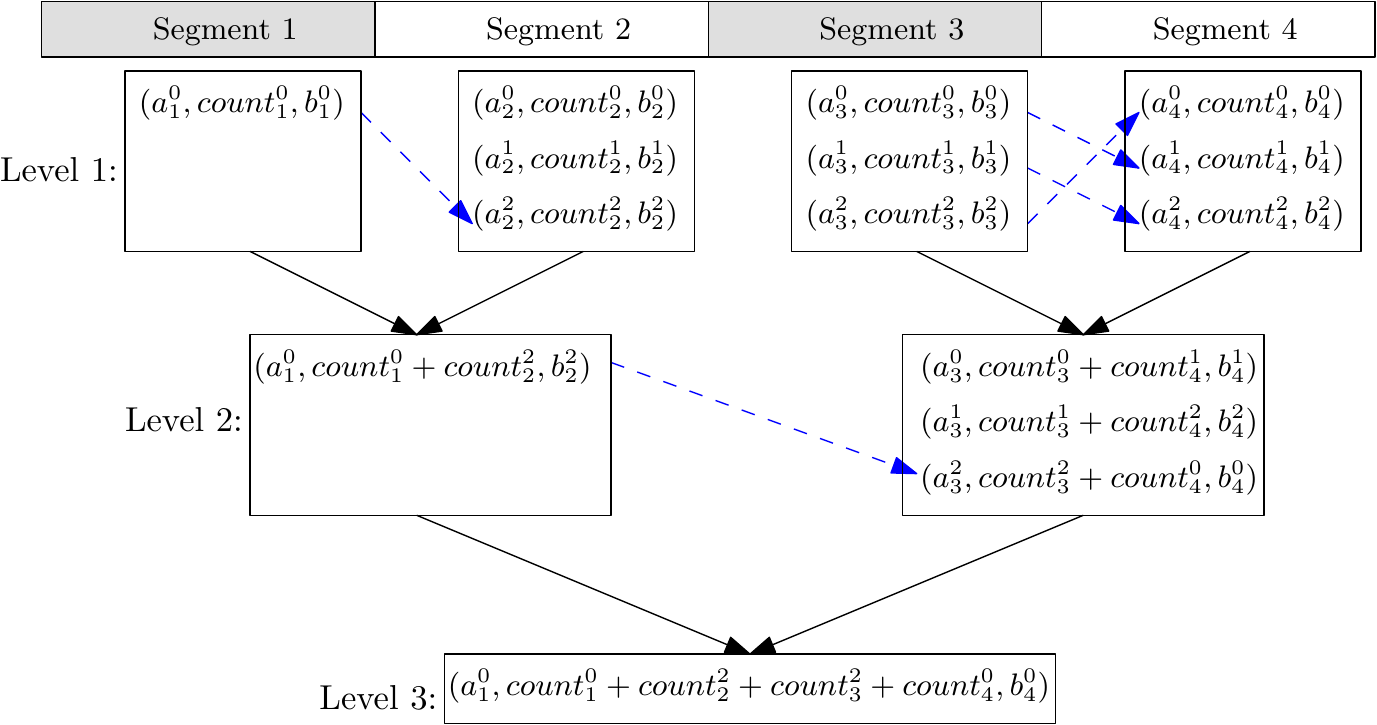}
	\caption{Illustration of a {\em Concatenate} step.}
	\label{fig:merge-step}
\end{figure}

The result of the
{\em Map} step for each segment $p$ is tuples of $(a_p^k, count_p^k, b_p^k)$. 
Based on these results, we design our {\em Concatenate} step for pairs of consecutive segments as follows, and illustrated in Figure~\ref{fig:merge-step}.


\begin{enumerate}
\item Start {\em Concatenate} step at level 1.
\item For level $i$, concatenate the tuples of segment $(j-1)2^i+1$ with the tuples of segment $(j-1)2^i+1+2^{j-1}$ for all possible $j$. The procedure for concatenate the tuples of $s^{th}$ segment and $t^{th}$segment is: find all pairs of tuples  $(a_s^k, count_s^k, b_s^k)$ and $(a_t^l, count_t^l, b_t^k)$ such that $b_s^k = a_t^l$, and then concatenate these pairs to obtain the next level ($i+1$) tuples  $(a_s^k, count_s^k + count_t^l, b_t^l)$ for $s^{th}$segment.
\item After all adjacent segment pairs are concatenated for level $i$, increase the level to $i+1$ and repeat the previous step until there is only one segment left for this level.
\end{enumerate}

It is worth mentioning that, at level $i$ of {\em Concatenate} step, segment $(j-1)2^i+1$ and segment $(j-1)2^i+1+2^{j-1}$ are considered as adjacent segments. We also choose segment number $p$ to be a power of 2, say $2^q$, so that the {\em Concatenate} step takes exactly $q+1$ levels and $2^{q+1}-1$ concatenate operations to finish.

\subsubsection{A Hybrid Algorithm}
\label{sec:hybrid}
\mc can greatly out-perform the PTPE algorithm in cases
when the number of episodes is small and some GPU cores are idle for the
PTPE algorithm. For other cases, the PTPE algorithm would run
much faster than \mc. So, as a practical approach, 
we would like to use their selective superiorities to automatically decide
the right algorithm to execute.
The logic for making these choices can be very simple: if the
GPU can be fully utilized with the PTPE algorithm, then we choose
it, else we choose \mc. In NVIDIA's CUDA framework, the GPU is fully utilized when:  
\begin{equation}
S > MP \times B_{MP} \times T_B
\label{eq:gpu-util}
\end{equation}
where $S$ is the number of episodes to be counted, $T_B$ is the number of threads per block as defined by the algorithm, $MP$ is the number of multiprocessors on the GPGPU, and $B_{MP}$ is the number of blocks that the compiler determines can be fit into one multi-processor.

However, another key factor that determines the selective superiority is
the size/length ($N$) of the episode being counted. For example, \mc will run 
slower on larger $N$, since the {\em Map} step needs $N$ state machines to count each 
event segment, forcing the {\em Concatenate} step to take more time to concatenate the 
state machines together. The performance of the
PTPE algorithm is also dependent on $N$, but will not change as much 
as \mc, since it only uses one state machine to count each episode. Therefore, we 
need to consider the effect of $N$ when deciding which algorithm to use: 
\begin{equation}
S > MP \times B_{MP} \times T_B \times f(N)
\label{eq:crossover}
\end{equation}
\noindent
where $f(N)$ is a performance penalty factor dependent on the episode length/level of the mining algorithm.
We define the cross-over point as the number of episodes beyond which we estimate that
the PTPE algorithm will outperform \mc.

By considering both GPU utilization and episode size, we propose the
{\em Hybrid} algorithm (A1) for mining temporal episodes on the GPU as shown in algorithm \ref{alg:hybrid}.
\begin{algorithm}[t]                      
\caption{GPU Mining Algorithm (A1)}          
\label{alg:hybrid}                           
\begin{algorithmic}[1]
\IF {$S > MP \times B_{MP} \times T_B \times f(N) $}
	\STATE Call PTPE Algorithm
\ELSE
	\STATE Call \mc Algorithm 
\ENDIF
\end{algorithmic}
\end{algorithm}

\subsection{A Two-Pass Elimination Approach: A2+A1}
\label{sec:a2plusa1}

After a thorough analysis of GPU performance of algorithm $A1$, we find the performance is largely limited by the requirement of large amount of shared memory and large number of GPU registers for each GPU thread. For example, if the episode size is 5, each thread requires 220 bytes of shared memory and 97 bytes of register file. It means that only 32 threads can be allocated on a GPU multi-processor, which has 16K bytes of shared memory and register file. When each thread requires more resources, only fewer threads can run on GPU at the same time, resulting in more execution time for each thread.

To address this problem, the only way is to reduce the complexity of the algorithm without losing correctness.
In this section, we introduce a two-pass elimination approach that more efficiently searches larger numbers of episodes,
further improving the overall performance. The idea is to use a far less complex algorithm, $A2$, to eliminate most of non-supported episodes, and only use the complex $A1$ algorithm to determine if the rest of episode is supported or not.
In order to introduce algorithm $A2$, we consider the solution 
to a slight relaxed problem, which plays an important role in our two-pass elimination approach.

\subsubsection{Less-Constrained Mining: Algorithm A2}
Let us consider a constrained version of 
Problem 1. Instead of enforcing both lower-limits and upper-limits on inter-event constraints, we 
design a counting solution that enforces only upper limits.

Let $\alpha'$ be an episode with the same event types as in $\alpha$, where $\alpha$ uses the original
episode definition from Problem 1.
The lower bounds on the inter-event constraints in $\alpha$ are relaxed for $\alpha'$ as shown below.
\[
\alpha' = \langle E_{(1)}^{\underrightarrow{(0,t_{high}^{(1)}]}}
					E_{(2)}
	        	\ldots^{\underrightarrow{(0,t_{high}^{(N-1)}]}}
	        E_{(N)}
	\rangle
\]

\begin{observation}
\label{thm:A2}
In Algorithm~\ref{alg:A1}, if lower-bounds of inter-event constraints in episode $\alpha$ are relaxed as $\alpha'$, the list size of $s[k], 1 \leq k \leq N$ can be reduced to 1.
\end{observation}

\begin{proof}

In Algorithm~\ref{alg:A1}, when an event of type $E_{(k)}$ is seen at time $t$ while going down the event sequence, $s[E_{(k-1)}]$ is looked up for at least one $t^{k-1}_{i}$, such that $t - t^{k-1}_{i} \in (0,t_{high}^{(k-1)}]$. Note that $t^{k-1}_{i}$ represents the $i^{th}$ entry of $s[E_{(k-1)}]$ corresponding the $(k-1)^{th}$ event-type in $\alpha$.

Let $s[E_{(k-1)}] = \{t^{k-1}_{1}\ldots t^{k-1}_{m}\}$ and $t^{k-1}_{i}$ be the first entry which satisfies the inter-event constraint $(0,t_{high}^{(k-1)}]$, i.e.,
\begin{equation}
0 < t - t^{k-1}_{i} \leq t_{high}^{(k-1)}
\label{eq:t1}
\end{equation}
Also Equation \ref{eq:t2} below
follows from the fact that $t^{k-1}_{i}$ is the first entry in $s[E_{(k-1)}]$ matching the time constraint.
\begin{equation}
t^{k-1}_{i} < t^{k-1}_{j} \leq t, \forall j \in \{i+1 \ldots m\}
\label{eq:t2}
\end{equation}
From Equation \ref{eq:t1} and \ref{eq:t2}, Equation \ref{eq:t3} follows.
\begin{equation}
0 < t - t^{k-1}_{j} \leq t_{high}^{(k-1)}, \forall j \in \{i+1 \ldots m\}
\label{eq:t3}
\end{equation}
This shows that every entry in $s[E_{(k-1)}]$ following $t^{k-1}_{i}$ also satisfies the inter-event constraint. This follows from the relaxation of the lower-bound. Therefore it is sufficient to keep only the latest time stamp $t^{k-1}_{m}$ only in $s[E_{(k-1)}]$ since it can serve the purpose for itself and all entries above/before it, thus reducing $s[E_{(k-1)}]$ to a single time stamp rather than a list (as in Algorithm \ref{alg:A1}). 
\end{proof}


\begin{algorithm}[t]
\begin{algorithmic}[1]
\REQUIRE Candidate episode $\alpha=\langle E_{(1)}\stackrel{(0,t_{high}^{(1)}]}{\longrightarrow}\ldots E_{(N)}\rangle$ is a $N$-node episode, event sequence $\textsl{S}=\{(E_i, t_i)\}, i \in \{1\ldots{n}\}$.
\ENSURE Count of non-overlapped occurrences of $\alpha$
\STATE $count = 0$; $s=[]$ //List of $|\alpha|$ time stamps
\FORALL{$(E,t) \in \textsl{S}$}
	\FOR{$i = |\alpha| \mbox{ to } 1$} \label{line:outer2}
		\STATE $E_{(i)} = i^{th}$ event type $\in \alpha$
		\IF{$E = E_{(i)}$}
			\STATE $i_{prev} = i - 1$
			\IF{$i > 1$}
				\IF{$t - s[i_{prev}] \leq t_{high}^{(i_{prev})}$}
					\IF{$i = |\alpha|$}
						\STATE $count ++$; $s=[]$; \textbf{break} Line: \ref{line:outer2}
					\ELSE
						\STATE $s[i]=t$
					\ENDIF
				\ENDIF
			\ELSE 
				\STATE $s[i]=t$
			\ENDIF
		\ENDIF
	\ENDFOR
\ENDFOR
\STATE \textbf{Output}: count
\end{algorithmic}
\caption{Less-Constrained Mining: A2}
\label{alg:A2}
\end{algorithm}

\subsubsection{Combined Algorithm: A2+A1}

Now, we can return to the original mining problem (with both upper and lower bounds). By combining Algorithm $A2$ with Algorithm $A1$, we can develop a two-pass elimination approach ($A2+A1$) that can 
deal with the cases on which Algorithm $A1$ cannot be executed.  Algorithm $A2+A1$ is as follows.

\begin{algorithm}[H]
\begin{algorithmic}[1]
\STATE (First pass) For each episode $\alpha$, run A2 on its less-constrained counterpart, $\alpha'$.
\STATE Eliminate every episode $\alpha$, if $count(\alpha') < CTh$, where $CTh$ is the support count threshold.  
\STATE (Second Pass) Run A1 on each remaining episode, $\alpha$, with both inter-event constrains enforced. 
\end{algorithmic}
\caption{Combined Algorithm: A2+A1}
\label{alg:A2A1}
\end{algorithm}

Algorithm $A2+A1$ yields the correct solution for Problem 1. Although the set
of episodes mined under the less constrained version are not a superset of those
mined under the original problem definition,  we can show the following result:

\begin{theorem}
\label{thm:A2A1}
$count(\alpha') \geq count(\alpha)$, i.e.,
the count obtained from Algorithm A2 is an upper-bound on the count obtained from Algorithm A1.
\end{theorem}


\begin{proof}
Let $h$ be an occurrence of $\alpha$. Note that $h$ is a map from event types in $\alpha$ to events in the data sequence $S$. Let the time stamps for each event type in $h$ be $\{t^{(1)}\ldots t^{(k)}\}$. Since $h$ is an occurrence of $\alpha$, it follows that
\begin{equation}
t^{i}_{low} < t^{(i)} - t^{(i-1)} \leq t^{i}_{high}, \forall i \in \{1\ldots k-1\}
\label{eq:t4}
\end{equation}
Note that $t^{i}_{low} > 0$. The inequality in Equation~\ref{eq:t4} still holds after we replace $t^{i}_{low}$ with $0$ to get Eqn.\ref{eq:t5}.
\begin{equation}
0 < t^{(i)} - t^{(i-1)} \leq t^{i}_{high}, \forall i \in \{1\ldots k-1\}
\label{eq:t5}
\end{equation}
The above corresponds to
the relaxed inter-event constraint in $\alpha'$. Therefore every occurrence of $\alpha$ is also an occurrence of $\alpha'$ but the opposite may not be true. Hence we have
that $count(\alpha') \geq count(\alpha)$. 
\end{proof}


In our two-pass elimination approach, Algorithm $A2$ is less complex and runs faster than Algorithm $A1$, because it reduces the time complexity of the
inter-event constraint check from $O(|s[E_{(k-1)}]|)$ to $O(1)$. Therefore, the performance of Algorithm $A2+A1$ is significantly better than Algorithm $A1$ when the number of episodes is very large and the number of episodes culled in the first pass is also large as shown by our experimental results described next.

\section{Experimental Results}
\subsection{Datasets and Testbed}

\subsubsection{Dataset Description}
Our datasets are drawn from both mathematical models of spiking
neurons as well as real datasets gathered by Wagenar et
al.~\cite{potter} in their analysis of cortical cultures. Both these
sources of data are described in detail in~\cite{Patnaik2008}. The
mathematical model involves 26 neurons (event types) whose activity is
modeled via inhomogeneous Poisson processes. Each neuron has a basal
firing rate of 20Hz and two causal chains of connections---one short
and one long---are embedded in the data. This dataset ({\it Sym26})
involves 60 seconds with 50,000 events. The real datasets 
({\it 2-1-33, 2-1-34, 2-1-35}) observe dissociated cultures on
days 33, 34, and 35 from over five weeks of development.
The original goal of this study was to characterize bursty
behavior of neurons during development. 

\subsubsection{Testbed and Runtime Parameters}
We evaluated the performance of our GPU algorithms on a machine equipped with Intel Core 2 Quad 2.33 GHz and 4GB system memory. We used a NVIDIA GTX280 GPU, which has 240 processor cores with 1.3 GHz clock for each core, and 1GB of device memory.

There are two CUDA runtime parameters we need to determine for each execution on the
GPU: number of threads per block, $T$, and the total number of blocks. The second parameter is always calculated based on $T$ so that all required computation can be finished with $T$ threads per block and
within one CUDA kernel function call. 

Parameter $T$ is determined by the
algorithm and the size of the episode ($N$). For the
PTPE algorithm, we calculate the maximum number of threads per block at each $N$. The larger $N$ is, more shared memory is needed per thread. When $N=1$, we use 128 threads, and as $N$ increases, the maximum number of threads per block decrease due to the shared memory limit. When $N=6$, we cannot have more than 32 threads per block. For \mc, the event stream is segmented into a number ($R$) of sub-streams, as mentioned in Section~\ref{sec:MapConcat}. Recall that we need to create multiple threads to count all sub-streams independently and run multiple state machines, as shown in Figure~\ref{fig:map-step}. The number of threads for each block $T$ can be calculated as $T=R \times N$, since there are $R$ sub-streams and $N$ state machines. Again, we must limit the number of sub-streams to reduce the number of threads due to the shared memory limit affected by $N$. For the 
Algorithm A2, we generate as many 
threads as possible per block until shared 
memory usage reaches the hardware limit (16KB). In this case, $T$ is normally much larger 
than 32, since we do not have a strict memory requirement for the GPU for Algorithm A2.

\subsection{PTPE vs \mc}

\begin{figure}[ht]
     \centering
     \subfigure[Run Times at Different Episode Sizes.]{
          \label{fig:mm-vs-n-vs-h}
          \includegraphics[width=4.5in]{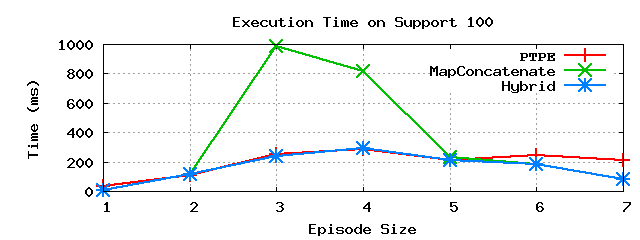}}
    \hspace{0.1in}
     \subfigure[Speedups at Different Support Thresholds.]{
          \label{fig:speedupsStream}
          \includegraphics[width=4.5in]{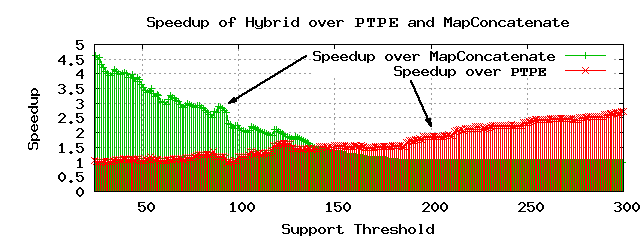}}
     \caption{Comparison of PTPE, \mc, and Hybrid on {\it Sym26} dataset.}
     \label{fig:a1}
\end{figure}

Our first set of tests evaluated the three algorithms:
PTPE, \mc, and 
the hybrid approach (A1). As seen in Figure~\ref{fig:mm-vs-n-vs-h}, it is clear that blindly choosing to execute the PTPE or \mc approach for \emph{all} levels is not the best solution. For episode sizes of 1, 2, and 5 both approaches complete in roughly the same amount of time. However, PTPE significantly outperforms \mc at episode sizes of 3 and 4 by 3.96X and 2.84X, respectively. On the other hand, PTPE performs slower than \mc for episodes of size 6 (1.32X) and 7 (2.63X).

These crossover points exist in all of our tests for this dataset (see supplementary information), and 
for lack of space, Table~\ref{tab:mm-vs-n} shows the crossover points determined experimentally 
for the Sym26 dataset. 

\begin{table}[h]
\caption{Crossover Points on number of episodes below which \mc should be run (for the
fewer episodes case). For other episode sizes---1, 2, and $>$8---\mc should be
chosen.}
\label{tab:mm-vs-n}
\begin{center}
\begin{tabular}{|c||c|c|c|c|c|c|}
\hline
Level & 3 & 4 & 5 & 6 & 7 & 8 \\
\hline
Crossover & 415 & 190 & 200 & 100 & 100 & 60\\
\hline
\end{tabular}
\end{center}
\end{table}%

\begin{figure}[ht]
     \centering
          \includegraphics[width=4.5in]{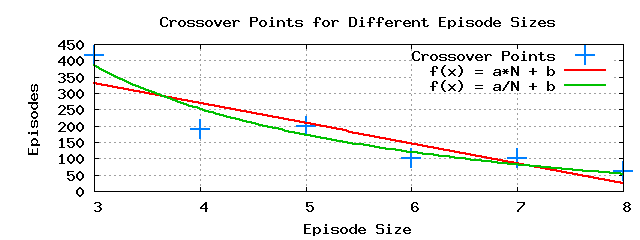}
     \caption{Crossover points fitted to $\frac{a}{N} + b$ and $a \times N + b$}
     \label{fig:fit}
\end{figure}

Recall Equation~\ref{eq:crossover} in Section~\ref{sec:hybrid} where optimal execution occurs when the GPU is fully utilized and a small factor of episode size is taken into account. Using the table above, with $M=30$, $T_B=32$, and $B_M=1$, we find that $f(N)$ of the form $\frac{a}{N} + b$ is a better fit than $a \times N + b$ as seen in Figure~\ref{fig:fit}.

With these crossover points determined, the hybrid approach was evaluated on the same support thresholds and the speedup for this approach over both PTPE and \mc is visible in Figure~\ref{fig:speedupsStream}. 
The range of improvement over PTPE is almost 3X and over 4.5X for \mc. When the number of episodes is large (i.e., low support threshold) there are enough episodes to fully utilize the GPU and as such Hybrid shows little improvement over PTPE. Conversely, Hybrid shows little improvement over \mc when the support threshold is high with very few episodes to search for.
%

\subsection{One-Pass (A1) vs Two-Pass (A2+A1)}

\begin{figure}[ht]
     \centering
     \subfigure[Execution time of Two-Pass and One-Pass algorithms for Support=3600 on Dataset {\it 2-1-35} at different episode sizes.]{
          \label{fig:culling-effect}
          \includegraphics[width=4.5in]{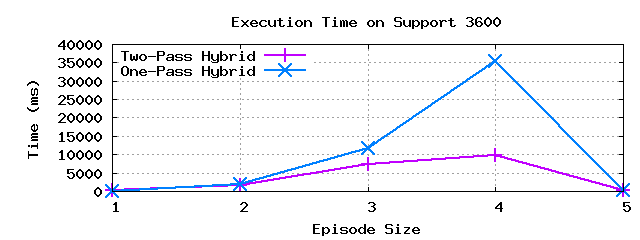}}
    \hspace{0.1in}
     \subfigure[Speedup of Two-Pass algorithm over One-Pass for multiple support thresholds on multiple datasets.]{
          \label{fig:culling-speedups}
          \includegraphics[width=4.5in]{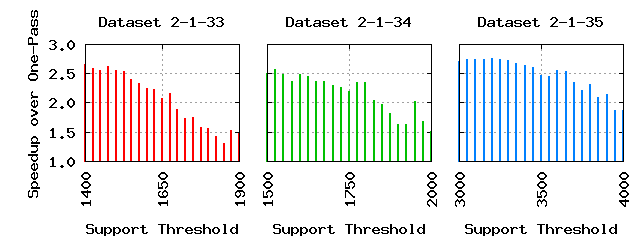}}
     \caption{Execution time and speedup comparison of One-Pass algorithm versus Two-Pass algorithm.}
     \label{fig:culling}
\end{figure}

As stated in Section~\ref{sec:a2plusa1}, the performance of algorithm $A1$ suffers from the requirement of large shared memory and large register file, especially when the episode size is big. So we introduce algorithm $A2$ that can eliminate most of the non-supported episodes and requires much less shared memory and register file,  then the complex algorithm $A1$ can be executed on much fewer number of episodes, resulting in performance gains.
The amount of elimination that $A2$ conducts can greatly affect the execution time at different episode sizes. In Figure~\ref{fig:culling-effect}, the two-pass algorithm eliminates over 99.9\% (43634 out of 43656) of the episodes of size four. The end result is a speedup of 3.6X over the hybrid $A1$ algorithm for this episode size and an overall speedup for this support threshold of 2.53X. Speedups for three different datasets at different support thresholds are shown in Figure~\ref{fig:culling-speedups} where in every case, the two-pass approach outperforms the one-pass approach with speedups ranging from 1.2X to 2.8X.

\begin{figure}[ht]
     \centering
          \includegraphics[height=1.4in]{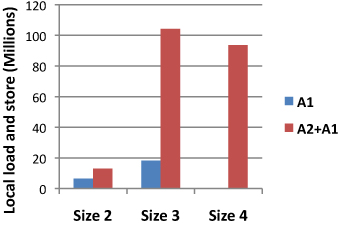} 
          \includegraphics[height=1.4in]{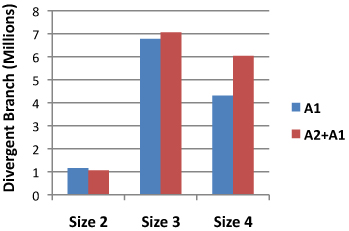} \\
          (a) \hspace{1.9in} (b) 
     \caption{Comparison between algorithm A1 and algorithm A2 for support threshold 1650 on dataset {\it 2-1-33}. (a) Total number of loads and stores of local memory. (b) Total number of divergent branches.   }
     \label{fig:twopass}
\end{figure}

We also use {\em CUDA Visual Profiler} to analyze the execution of algorithm $A1$ and $A2$ to give a quantitative measurement of how $A2$ out-performs $A1$ on the GPU. 
We have analyzed various GPU performance factors, such as GPU occupancy, coalesced global memory access, shared memory bank conflict, divergent branching, and local memory loads and stores. We find the last two factors are primarily attributed to the performance difference between $A1$ and $A2$, which are shown in Figure~\ref{fig:twopass}. Algorithm $A1$ requires 17 registers and 80 bytes of local memory for each counting thread, while algorithm $A2$ only requires 13 registers and no local memory. Since local memory is used as supplement for registers and mapped onto global memory space, it is accessed very frequently and has the same high memory latency as global memory. In Figure~\ref{fig:twopass} (a), the total amount of local memory access of both two-pass approach and one-pass approach comes from algorithm $A1$. Since algorithm $A2$ eliminates most of the non-supported episodes and requires no local memory access, the local memory access of two-pass approach is much less than one-pass approach when the size of episode increases. At the size of 4, $A2$ eliminates all episode candidates, thus there is no $A1$ execution and no local memory access, resulting a large performance gain for two-pass approach over one-pass. As shown in Figure~\ref{fig:twopass} (b), the amount of divergent branching also affects the GPU performance difference between the two-pass approach and the one-pass approach.    

\subsection{GPU vs. CPU}
To demonstrate the performance gain of our GPU approach, we implemented Algorithm~\ref{alg:A1} for a quad-core CPU with four threads, and tested on our testbed machine. The CPU implementation is written in C++ and optimized for sequentially executed applications. For example, since each CPU thread counts a large number of episodes, we can read the event stream exactly once for each thread, and update all state machines in that thread with each event. In addition, we used an acceleration structure to speed up the search for which the state machine needs to be updated. In addition, we also implement two-pass ($A2+A1$) elimination approach on CPU to get a fair performance comparison against our GPU implementation.

\begin{figure}[ht]
     \centering
          \includegraphics[width=5.0in]{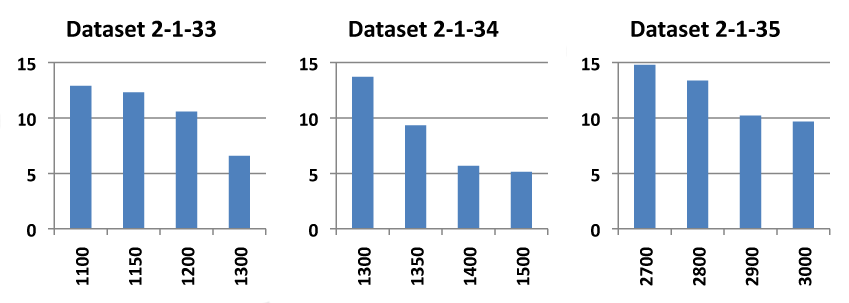}
     \caption{Speedup of GPU implementation over the CPU implementation}
     \label{fig:cpu-gpu}
\end{figure}

Compared with the performance of the CPU implementation, our GPU algorithms exhibit a significant speedup, as shown in Figure~\ref{fig:cpu-gpu}. For the 2-1-35 dataset, the speedup is approximately 15-fold for a support threshold of 2700.  

\subsection{Mining Evolving Neuronal Circuits}
Due to space limitations, we are unable to discuss the biological significance
of our mined episodes. In addition to understanding the activity of
an ensemble of neurons,
one of the advantages of our GPGPU solution is to mine evolving
cultures, and watch the progression of neuronal development in real-time.
Please see our supplemental website http://neural-code.cs.vt.edu/gpgpu
for videos depicting how our mining reveals key characteristics of
cortical culture growth.

%

\section{Discussion}
We have presented a powerful and non-trivial framework for conducting
frequent episode mining on GPGPUs and shown its capabilities for mining
neuronal circuits in spike train datasets. For the first time, neuroscientists
can enjoy the benefits of data mining algorithms without
needing access to costly and specialized clusters of workstations.
Our supplementary website (http://neural-code.\hskip0ex cs.\hskip0ex vt.\hskip0ex edu/\hskip0ex gpgpu) provides
auxiliary plots and videos demonstrating how we can
track evolving cultures to reveal 
the progression of neural
development in real-time.

Our future work is in four categories. First, our experiences with
the neuroscience application have
opened up the interesting topic of mapping finite state machine
based algorithms onto GPGPUs. A general framework to map any finite
state machine algorithm for counting will be extremely powerful not just
for neuroscience but for many other areas such as (massive) sequence analysis
in bioinformatics and linguistics. Second, the development of
{\it {\mc}} highlights the importance of
developing new, additional, programming abstractions specifically geared toward
data mining on GPGPUs. Third, we found that the two-pass approach performs significantly better than running the complex counting algorithm on the entire input. The first pass solves a simpler problem, and generates an upper bound of the solution to the original problem. This significantly reduces the input size for the complex second pass, speeding up the entire process. Finally, we wish to integrate more aspects of
the application context into our algorithmic pipeline, such as candidate
generation, streaming analysis, and rapid ``fast-forward'' and ``slow-play''
facilities for visualizing the development of neuronal circuits.



\end{document}